\documentstyle[12pt,psfig]{article}

\def\baselinestretch{1.2}
\def\href#1#2{#2}  

\newcommand{\norm}[1]{\raise.3ex\hbox{:} #1 \raise.3ex\hbox{:}\,}

\def\det{{\rm det}}

\newcommand{\beq}{\begin{equation}}
\newcommand{\eeq}{\end{equation}}
\newcommand{\al}{\alpha^{'}}

\newcommand{\nl}{\hspace{-.65cm}}

\def\appendix{{\newpage\section*{Appendix}}\let\appendix\section%
        {\setcounter{section}{0}
        \gdef\thesection{\Alph{section}}}\section}

\begin{document}

\begin{titlepage}

\begin{flushright}
NSF-ITP-99-085\\
hep-th/9907166
\end{flushright}
\vfil\vfil

\begin{center}

{\Large {\bf Non-Commutative Yang-Mills and \\The AdS/CFT
Correspondence}}

\vfil

\vspace{10mm}

Akikazu Hashimoto\\

\vspace{10mm}

Institute for Theoretical Physics\\ University of California,
Santa Barbara, CA 93106\\
aki@itp.ucsb.edu\\

\vspace{10mm}

N. Itzhaki\\

\vspace{10mm}

Department of Physics\\ University of California,
Santa Barbara, CA 93106\\
sunny@physics.ucsb.edu\\

\vfil
\end{center}

\begin{abstract}
\noindent We study the non-commutative supersymmetric Yang-Mills
theory at strong coupling using the AdS/CFT correspondence. The
supergravity description and the UV/IR relation confirms the
expectation that the non-commutativity affects the ultra-violet but
not the infra-red of the Yang-Mills dynamics.  We show that the
supergravity solution dual to the non-commutative ${\cal N}=4$ SYM in four
dimensions has no boundary and defines a minimal scale.  We also show
that the relation between the $B$ field and the scale of
non-commutativity is corrected at large coupling and determine its
dependence on the 't Hooft coupling $\lambda$.
\end{abstract}

\vfil\vfil\vfil
\begin{flushleft}
July 1999
\end{flushleft}

\end{titlepage}
\renewcommand{\baselinestretch}{1.05}  

Classical supersymmetric Yang-Mills theory (SYM) in
$(p+1)$-dimensional space can be generalized to SYM on non-commutative
spaces \cite{connes}.  Since the generalization involves infinitely
many higher order terms, it is very hard to provide a pure field
theory proof that such a theory is consistent at the quantum level.
String theory provides a way to obtain these theories by considering
the decoupling limit of D$(p-2)$-branes in type II string theories on
$T^2$ with a background NSNS 2-form field $B_{\mu \nu}$ polarized
along the plane of the torus \cite{CDS,DougHull}.  The fact that
non-commutative SYM is obtained from string theory, in a limit which
does not involve gravity, suggests (if string theory with constant $B$
field is consistent) that the non-commutative SYM, at least with
sixteen supercharges, is a consistent theory at the quantum level.

In this article, we take advantage of the fact that exactly the same
decoupling limit leads to the near horizon geometry of the D$p$-branes
\cite{mald,IMSY} to learn about the non-commutative SYM at large
coupling\footnote{Related ideas were discussed in
\cite{CDS,seibergtalk}.}.  We begin by reviewing the argument of
\cite{CDS,DougHull} regarding how the background $B$-field gives rise
to a non-commutativity of scale $\Delta$.  We describe a scaling limit
on the field theory side which keeps the $\Delta$ finite while sending
$\alpha'$ to zero to decouple the stringy excitations.  Then, we
consider the same scaling limit in the dual closed-string picture, and
find that the background $B$-field changes the dynamics of the closed
strings only in the region far away from the horizon (UV), while
keeping the dynamics near the horizon (IR) unaffected.  We find that
the background geometry does not have any boundary, which we interpret
as the manifestation of the fact that theories on non-commutative
spaces do not have a local UV description.

Following \cite{DougHull}, we consider D$(p-2)$-branes in a weakly
coupled type II string theory, oriented along the $x_0,...,x_{p-2}$
directions.  Consider compactifying $x_{p-1}$ and $x_p$ coordinates on
a square torus of radius $R$ and  turn on a constant NSNS
$B$-field polarized along the plane of this torus.  In the absence of
the $B$-field and in the limit 
\beq\label{rlimit}
\frac{R}{\al}=\frac{1}{\Sigma}={\rm fixed}, \qquad \al \rightarrow 0, \eeq
it is natural to describe this system in the T-dual picture of
D$p$-branes wrapping the dual torus whose size $\Sigma$ is
macroscopic.

As we shall see shortly, to obtain a finite non-commutativity
scale\footnote{In the estimate of scales, numerical factors of order
one are ignored.} in the decoupling limit, the $B$ field has to satisfy
\beq\label{blimit}
\Delta^2 = B\al = {\rm fixed}, \qquad \al \rightarrow 0.
\eeq
Thus the $B$ field has to be very large, and in the presence of such a
strong $B$-field, the T-duality is strongly modified.  Note that since
we would like to make contact with the AdS/CFT correspondence we are
using conventions which are natural from the supergravity point of
view. These conventions are different then the ones used in the recent
non-commutative geometry literature.  In our conventions $
S_B=\frac{1}{4\pi \al} \int d\sigma^2 \epsilon^{\alpha\beta}B_{\mu\nu}
\partial_{\alpha}x^{\mu} \partial_{\beta}x^{\nu}$ where $x$ has the
dimension of length (if $x$ is compactified then $x \sim x +2\pi R$)
and hence $B$ and $G$ are dimensionless.  The fact that a large $B$ field
strongly modifies the T-duality transformation can be seen from the
form of the transformation of the matrix $E$ \cite{GPR} which in our
notations takes the following form:
\beq\label{7} E=\frac{R^2}{\al}(G+B)=\frac{R^2}{\al} \left(
\begin{array}{cc} 1 & B \\ -B & 1
\end{array}\right).
\eeq
 T-duality takes $E$ to $E^{-1}$ and
so the dual radius in the presence of $B$ field is \beq\label{3}
\Sigma_B =\Sigma \frac{\al}{\Delta^2} \eeq
which is not macroscopic, as $\Sigma_B$ vanishes when $\al \rightarrow
0$. Therefore, in the presence of a large $B$ field, we cannot
T-dualize to end up with D$p$-branes wrapping $p+1$ macroscopic
dimensions. Instead we end up with only $p-1$ macroscopic dimensions.

On the other hand, due to the Dirichlet boundary conditions along the
$x_{p-1}$ and the $x_p$ directions, the energy
\beq\label{1} E_s= \frac{1}{\Sigma} \eeq
of a string stretched between the images of the branes is not affected
by the $B$ field.  Therefore, from the point of view of the open
strings living on the D$(p-2)$-branes, the theory has $p+1$
macroscopic dimensions.

We have concluded that although the D-branes are wrapping only $p-1$
large directions, the field theory living on the branes knows about
$p+1$ large dimensions!  This apparent mismatch of the number of
macroscopic dimensions is disturbing in light of the AdS/CFT
conjecture which implies a duality between the closed string and open
string description of the branes. The goal of this article is to
resolve this discrepancy and to provide an interpretation of the
non-locality in the dual closed string picture. To achieve this goal,
it is useful to first examine the relation between the $B$-field and
the non-commutativity scale $\Delta$ more closely \cite{DougHull}.

Instead of T-dualizing twice, Douglas and Hull considered the
following chain of ``duality'' transformations. First, they perform a
T-duality along one of the cycles. Due to the presence of the
$B$-field in the background, the dual torus will not be
rectangular. The D$(p-2)$-branes have now become D$(p-1)$-branes and
the light degrees of freedom of equation (\ref{1}) are now momentum
modes along the D$(p-1)$-branes and the winding modes along the short
cycle.
\begin{figure}
\centerline{\psfig{file=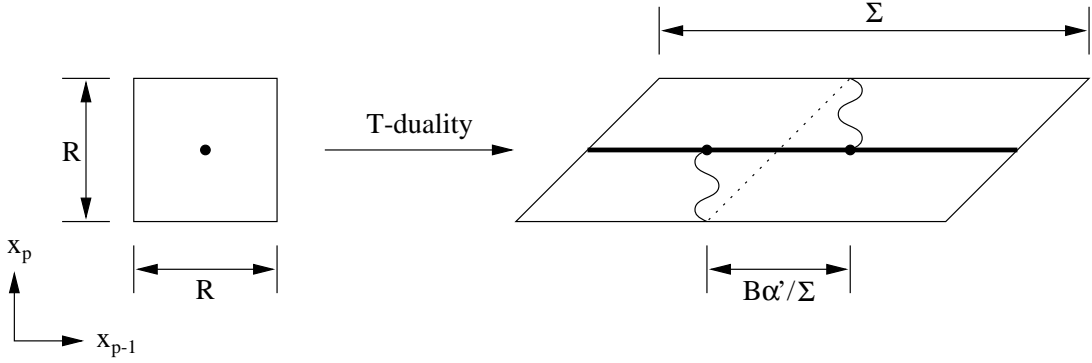}}
\caption{D$(p-2)$-branes in square torus of radius $R \ll
\sqrt{\alpha'}$ and B-field flux and its T-dual D$(p-1)$-branes on a
skewed torus. Skewed geometry of the torus gives rise to non-locality
in the open string excitations living on the
D$(p-1)$-brane.\label{figa}}
\end{figure}
To minimize the energy, winding modes will wind the torus in the
shortest path illustrated in figure \ref{figa}. Therefore, both the
momentum and the winding modes have masses of order $1/\Sigma $. Due
to the skewed shape of the dual torus, however, the winding modes are
delocalized along the D$(p-1)$-brane world volume by length of order
\beq\label{2}
\Delta_{p-1} =\frac{B  \al w}{\Sigma},
\eeq
where $w$ is the winding number.  The Compton wavelength along the
$x_{p}$ direction associated with such a state is of order $\Delta_{p}
= \Sigma/w$.  Combining these results, the scale of non-locality
comes out to
\beq\label{17}
 \Delta^2 \equiv \Delta_{p-1} \Delta_{p} = B \al .  \eeq
Readers are referred to \cite{DougHull,MLi} for more details.

Since the $B$ field has a finite effect on the field theory living on
the brane in the limit (\ref{rlimit}) and (\ref{blimit}), it should
also have a finite effect on the dual supergravity description of the
theory.  At first sight this does not seem to be the case, since on
the closed string side of the duality we do not end up with $p+1$
large dimensions.  The resolution stems from taking proper account of
the effect of the D-brane background geometry in the near horizon
region. For concreteness, let us concentrate on the conformal case by
setting $p=3$. Our conclusions can be generalized immediately to the
non-conformal cases with $p \ne 3$.  The string frame solution in the
presence of D1-branes and their images coming from the $T^2$
compactification is \cite{HS}\footnote{The solution is not modified by
a constant $B$-field since $H=dB=0$ and does not act as a source for
the other supergravity fields.}
\begin{eqnarray}
&& ds^2=f^{-1/2}(-dt^2 +dx_1^2)+
f^{1/2}(dx^2_2+...+dx_9^2),
\nonumber\\
&& e^{\phi}= f^{1/2},\\
&& B_{23}=\frac{\Delta^2}{\al}, \nonumber  
\end{eqnarray}
where $f$ is the harmonic function of the transverse coordinates
$U=\sqrt{x_4^2+...+x_9^2}/\al$ and we Poisson re-sum over the images
coming from the $T^2$ directions $x_2$ and $x_3$, \beq f=1+
\frac{\lambda}{\al {}^2 U^4}.  \eeq The 't Hooft coupling constant of
the four dimensional field theory is denoted by $\lambda = 2 g_{YM}^2
N$.  We see that in the near horizon region the longitudinal
directions shrinks while the transverse directions blow up.  Proper
treatment of this effect amounts to setting $g_{22}$ and $g_{33}$
equal to $f^{1/2}$ instead of $1$ in equation (\ref{7}).  Therefore,
in the near horizon limit, $g_{22}$ and $g_{33}$ blow up just at the
right strength to compete with the effect of the $B$ field.  Applying
T-duality to this background\footnote{Under T-duality, the dilaton
transforms according to $\phi' = \phi - {1 \over 4} \log\left( \det{g}
/ \det{g'} \right)$ \cite{GPR}.}and taking the field theory decoupling
limit, we obtain\footnote{This background can also be found by
applying the decoupling limit to equation (2.20) in
\cite{russotseyt}.}
\begin{eqnarray}\label{9} 
&& ds^2= \al \left\{
\frac{U^2}{\sqrt{\lambda}}(-dt^2+dx_1^2) +\frac{\sqrt{\lambda}
U^2}{\lambda + U^4 \Delta^4}(dx_2^2+dx_3^2)
+\frac{\sqrt{\lambda}}{U^2}dU^2 +\sqrt{\lambda}d \Omega_5^2 \right\},
\nonumber \\ 
&& e^{\phi}=\frac{\lambda}{4\pi N}
\sqrt{{\lambda \over \lambda + \Delta^4 U^4}},\\  
&& B_{23} = -{\alpha' \Delta^2 U^4 \over \lambda + \Delta^4 U^4},\nonumber
\end{eqnarray}
with periodicities $x_2 \sim x_2 +2 \pi \Sigma$ and $x_3 \sim x_3 +2 \pi \Sigma$.
To avoid the finite size effects we take $\Sigma \gg \Delta $.  In the
spirit of \cite{mald} we conjecture that the type IIB string theory on 
this background is dual to non-commutative SYM with non-commutative
$x_2$-$x_3$ plane.

Equation (\ref{9}) is the main result of this paper. It describes the
dual supergravity background corresponding to the same scaling limit
used to define SYM on non-commutative geometries. Let us pause and
make a few comments on the qualitative features of (\ref{9}).

\nl $\bullet$ The geometry (\ref{9}) is the effective description of
our system when the curvature and the coupling are small. According to
(\ref{9}), the dilaton is small everywhere in the large $N$ limit.
Unlike in AdS the invariant curvature in string units depends on $U$.
However, it is always of the order of the AdS curvature,
$1/\sqrt{\lambda}$.  Thus for large 't Hooft coupling we can trust the
solution everywhere.  Notice that after the T-duality the $B$ field is
not a constant and hence $H\neq 0$.

\nl $\bullet$ The observation of \cite{mald} that $U$ plays the role
of energy scale on the field theory side is not modified by $\Delta$
as the energy of a string stretched between the collection of the
branes and a probe brane is the same as in the AdS case.  This follows
from the fact that $\Delta$ does not modify the relation $\sqrt{g_{tt}
g_{UU}} =\al $ which determines the energy of the string.
Alternatively, in the D1-branes language, before the T-duality, the
presence of the $B$ field does not modify the energy of the open
strings.

\nl $\bullet$ The isometries of equation (\ref{9}) are $SO(1,1)\times
SO(2) \times SO(6)$.  $SO(1,1)\times SO(2)$ and the translation
invariance are the remnants of the $SO(4,2)$ of $AdS_5$.  The fact
that the conformal and special conformal transformation are broken
follows from the presence of the scale $\Delta$.  The fact that
Lorentz invariance is broken to $SO(1,1)\times SO(2)$ agrees with the
effect of equation (\ref{17}) on the field theory side. The $SO(6)$ is
the isometry of the 5-sphere, and corresponds to the $SU(4)$
R-symmetries of the ${\cal N}=4$ supersymmetry algebra.  The fact
$SU(4)$ is not broken by the non-commutativity implies that the
supersymmetry is not broken by the non-commutativity either.  Note
that since the conformal invariance is broken the number of
supercharges is 16 and not 32. Furthermore, the background is not
self-dual with respect to S-duality.

\nl $\bullet$ The presence of a finite non-locality scale implies that
the dynamics at large distances (compared to the non-locality scale)
is not affected while the short distances dynamics is drastically
changed.  This is exactly what we see in the supergravity description.
Equation (\ref{9}) describes the usual $AdS_5 \times S^5$ solution
with a constant dilaton in the IR $(U \rightarrow 0)$, while the
solution is strongly modified in the UV.  On the supergravity side the
non-commutativity scale can be read off from the point at which the
modification to the $AdS_5 \times S_5$ background becomes of order
one.  This happens at
\beq\label{aa} U=\frac{\lambda^{1/4}}{\Delta}.  \eeq
Using the UV/IR relation \cite{SW,peetpol}, $L\sim \sqrt{\lambda}/U$, we find
that at large 't Hooft coupling the non-commutativity scale is not
$\Delta =B \al$ but rather\footnote{The general expression for
arbitrary $p$ is $\tilde\Delta = \Delta^{2 (5-p)/ (7-p)}
\lambda^{1/(7-p)}.$}
\beq\label{b}
\label{11} \tilde{\Delta} = \Delta \lambda ^{1/4}= B \al  \lambda ^{1/4}.
\eeq
The fact that the $\tilde{\Delta}$ is different than $\Delta$ is an
indication that the relation between $B$ and the non-commutativity
scale receives quantum corrections. It would be interesting
to study the corrections in perturbation theory.

Although the discussion above provides some evidence that the theory
acquires a new dynamical scale at $\lambda^{1/4} \Delta$, we have not
yet demonstrated (other than by construction) that this scale is
associated with non-commutative geometry. Non-commutative geometry has
a built in minimal distance scale, and we would like to see this from
the supergravity point of view.  A clean way to see that there is such
a minimal distance is the following.  Let us add a non-commutativity
scale in the $x_0$-$x_1$ plane (by starting with type IIB D-instantons
in the presence of $B_{23}$ and $B_{01}$). The corresponding
supergravity solution (in the Euclidean space) is
\begin{eqnarray}\label{l} ds^2&=& \al \left\{
\frac{\sqrt{\lambda}
U^2}{\lambda + U^4 \Delta_{01}^4}(dx_0^2+dx_1^2) +\frac{\sqrt{\lambda}
U^2}{\lambda + U^4 \Delta_{23}^4}(dx_2^2+dx_3^2)
+\frac{\sqrt{\lambda}}{U^2}dU^2 +\sqrt{\lambda}d \Omega_5^2 \right\}, \nonumber \\
e^\phi & = & {\lambda \over 4 \pi N} 
\sqrt{ {\lambda \over \lambda+\Delta_{01}^4 U^4}} 
\sqrt{ {\lambda \over \lambda+\Delta_{23}^4 U^4}}  .
\end{eqnarray}
To simplify the discussion, let us set $\Delta_{01} = \Delta_{23} =
\Delta$\footnote{This case is the one relevant for
\cite{nikita}.}. This geometry is manifestly invariant under the
transformation
\beq \label{UUtilde} \tilde{U}=\frac{\lambda^{1/2}}{U \Delta^2}.  \eeq
As we mentioned earlier, this geometry asymptotes to $AdS_5 \times
S_5$ in the small $U$ (IR) limit.  Thus from the IR point of view the
boundary should be at $U=\infty$.  However, as a result of the $U
\leftrightarrow \tilde{U}$ invariance, the $U \rightarrow \infty$
limit is also $AdS_5 \times S_5$.  So from the UV point of view the
boundary should be at $\tilde{U}=\infty$ and therefore at $U=0$.  In
fact the space described by equation (\ref{l}) is essentially two AdS spaces
which are glued together in such a way that geodesics starting from
the region near the horizon of one of the AdS spaces reach the
interior of the other AdS, and so this space has no
boundary\footnote{The background (\ref{9}), corresponding to the case
of vanishing $\Delta_{01}$, does have a boundary at $U = \infty$ but
with only two dimensions parameterized by $t$ and $x_1$.}.  Now, a
local field theory is defined at short distances, and in terms of the
AdS/CFT correspondence this means that the microscopic structure of
the theory is encoded on the boundary of the AdS space.  Having a
non-commutative theory would imply that we should not be able to
define the theory at short distances. The fact that our geometry has
no boundary is the supergravity manifestation of this fact, and the
minimal distance scale is set by the self-dual.

We should stress that (\ref{UUtilde}) is not a duality.  Only the
string frame metric is invariant under this
transformation\footnote{Both the dilaton and the $B$-field are not
invariant with respect to (\ref{UUtilde}).}. The same is not true for
the metric in the Einstein frame because of the non-trivial dilaton
background.  In fact, the Einstein frame metric asymptotes to ten
dimensional Minkowski spacetime at large $U$. This might be a useful
observation in attempts to understand flat space-time holography.
Despite the fact that (\ref{UUtilde}) is not a duality, it resembles a
similar relation in T-duality. Perhaps this analogy will prove useful
for the future investigations of non-commutative SYM.

The goal of this investigation was to understand the mechanism of
non-locality in the non-commutative SYM at large gauge coupling from
the dual supergravity description. We were guided by the intuition
that when the effect of non-locality of order $\Delta$ in SYM is
turned on, the dynamics at length scales longer than $\Delta$ is
unaffected, whereas the dynamics at length scales smaller than
$\Delta$ is drastically changed. We formulated a scaling limit of open
string dynamics which keeps $\Delta$ fixed while sending $\alpha'$ to
zero.  Following the same scaling limit and applying T-duality, we
obtained the background geometry (\ref{9}). In the $U \rightarrow 0$
limit, this geometry asymptotes to the usual $AdS_5 \times S_5$
geometry, confirming our expectation that the IR dynamics is
unaffected by non-commutativity. As $U$ is increased, the geometry
starts to deviate from its AdS limit. Because of the $U
\leftrightarrow \tilde{U}$ invariance, the geometry in the $U
\rightarrow \infty$ limit is also an $AdS_5 \times S_5$ geometry,
whose natural radial coordinate is $\tilde{U}$.  We find therefore
that the supergravity dual to the non-commutative SYM does not have a
boundary, unlike the previously encountered examples of the
boundary/bulk correspondence.  Although this might seem surprising at
first sight, it follows quite naturally from the fact that field
theories on non-commutative spaces do not admit a local UV
description.  The scale of non-commutativity $\tilde{\Delta} =
\lambda^{1/4} \Delta$ can be read off from the self-dual scale of the
$U \leftrightarrow \tilde{U}$ transformation. This non-commutativity
scale disagrees with the non-commutativity scale computed in the
weakly coupled limit in \cite{DougHull} by a certain functional
dependence on the 't Hooft coupling constant $\lambda$. We interpret
this to mean that the relation between $\Delta$ and the background $B$
field receives quantum corrections.

\section*{Note Added}

We have learned that J. Maldacena and J. Russo have  considered related 
issues \cite{juan}.

\section*{Acknowledgments}

NI would like to thank S. Yankielowicz for discussions.  The work of AH is
supported in part by the National Science Foundation under Grant
No. PHY94-07194. The work of NI is supported in part by NSF grant
No. PHY9722022.

\begingroup\raggedright\endgroup

\end{document}